\documentclass{ws-ijmpa}

\begin{document}

%%%%%%%%%%%%%%%%%%%%% Publisher's Area please ignore %%%%%%%%%%%%%%%
%
\catchline{}{}{}{}{}
%
%%%%%%%%%%%%%%%%%%%%%%%%%%%%%%%%%%%%%%%%%%%%%%%%%%%%%%%%%%%%%%%%%%%%

\title{Describing non-$q\bar q$ candidates.}

\author{J. Vijande, F. Fern\'andez, A. Valcarce}

\address{Nuclear Physics Group,\\
University of Salamanca,\\
Plaza de la Merced s/n, E-37008 Salamanca, Spain}

\maketitle

\pub{Received (Day Month Year)}{Revised (Day Month Year)}

\begin{abstract}
Despite the apparent simplicity of meson spectroscopy there are some states
which cannot be accommodated in the usual $q\bar q$ structure. Among
them there are either exotic states as the $X(1600)$ or the recently measured
charm states $D_{sJ}^*$ and $X(3872)$ and some of the light scalar mesons. 
In this work we present a possible description of these states in terms
of tetraquarks.
\end{abstract}

\vspace{0.4cm}
The existence of low-energy multiquark states cannot be discarded
from the quark model point of view. Gauging a free quark theory through the
$SU(3)_c$ group all $(q\bar q)^n\,(qqq)^m$ states, being $n$ and $m$ integers, are
allowed. Although suggested long time ago\cite{jaff},
only recently experimental evidence of the existence of these states has been 
obtained. Apart from the widely discussed pentaquark ($n=m=1$) during the last
years several tetraquark candidates ($n=2,m=0$) have been suggested. Among them,
there are states compatible with meson quantum numbers, as it is the case of the 
$D^*_{sJ}$'s and the $X(3872)$\cite{barn}. However other structures have to be clearly 
ascribed to a multiquark state as for example the $X(1600)$, being an isospin
two system\cite{lili}.

The study of four quark systems has been done in two different directions.
There have been some theoretical works specifically devoted to a particular set of states\cite{tere},
while others did a more general study but in any case making a detailed comparison 
with $q\bar q$ predictions within the same model\cite{zuzu}. 
The exciting scenario created by the new data obtained at BaBar, CLEO, FOCUS
and Belle claims for a comprehensive study where two- and four-quark states are
simultaneously addressed.

The $q-q$ interaction used in this work has been derived from a complete 
study of the meson spectra from the light pseudoscalars to bottomonium\cite{vij2}, 
compatible with the description of $NN$ data and the baryon spectra.
To solve the four body problem we perform a variational calculation considering
non-quadratic terms in the radial wave function that were neglected in previous works\cite{epja}.
These terms, which play a minor role in the description of the light-heavy tetraquarks\cite{epja}, 
have an important influence in the $(qq)(\bar q\bar q)$ and $(qs)(\bar q\bar s)$ tetraquarks, 
where $q$ stands for $u$ and $d$ quarks, and must be included
to obtain a reliable description of these states.

\section{$X(1600)$}
This state has been reported with a mass of $1600\pm100$ MeV\cite{pdg}. It
has been observed in the reaction $\gamma\gamma\rightarrow\rho\rho$ near
threshold with quantum numbers $I^GJ^{PC}=2^+(2^{++})$\cite{albr}. This implies
that it cannot be described as a $q\bar q$ state, being therefore an exotic meson.
Its quantum numbers can be easily
obtained as a tetraquark made of four light quarks coupled to $I=2$,
$S=2$ and $L=0$. Our model predicts for this configuration an energy of $1544$ MeV,
in excellent agreement with the experimental data.
Let us emphasize that in the tetraquark calculation there are no free parameters, all of them being
fixed in the $NN$ interaction and hadron spectroscopy\cite{vij2}.

\section{The charm sector: $D_{sJ}^*(2317)$ and $X(3872)$}
BaBar has reported a narrow state near 2317 MeV known as $D_{sJ}^{\ast }(2317)$\cite{baba}
with quantum numbers $J^P=0^+$. Its identification with a conventional $c\bar s$ quark state 
appears not possible due to its low mass\cite{barn}. Our results for the scalar $c\bar s$ member of the $P-$wave triplet is 2470 MeV, 
too heavy to be identified with the $D_{sJ}^{\ast }(2317)$.

This state has been observed in a strong or electromagnetic decay to $D_s^+
\pi^0$ so it must at least contains $c$ and $\bar s$ quarks. The most obvious
possibility of a tetraquark will be $(qc)(\bar q \bar s)$ coupled to $I=0$ or $1$.
We have obtained for this configuration 2449 MeV for the isovector case and 2503 MeV 
for the isoscalar state. 
Although they are still too heavy to be identified with the $D_{sJ}^*(2317)$,
different alternatives could improve the situation. The first one is a possible $\bar s
c\leftrightarrow qc\bar q \bar s$ coupling. In this case if the mixing is fitted to
reproduce the mass of the $D_{sJ}^{\ast }(2317)$ we obtain a 55\%
$c\bar s$ component and another isoscalar state with a mass of 2655 MeV, 
compatible with the recently discovered $D_{sJ}^{\ast }(2623)$ state at
SELEX\cite{sele}.
A different approach was proposed by Barnes {\it et al} \cite{barn} considering 
a possible isospin mixing. In our
case we can fit this mixing to obtain the experimental energy, predicting an 
isospin symmetry breaking (58.5\% for the I=1 component) and an orthogonal
state with a mass of 2635 MeV. Although its mass is also close to the one found
by SELEX this strong isospin mixing is difficult to be justified within a $q-q$
interaction. Another possibility would be the influence of
three-body color forces\cite{dmit}, once included one
can obtain the mass of the $D_{sJ}^{\ast }(2317)$ fitting its strength.

The most recent of the states discovered in the charm sector is the $X(3872)$, 
which was reported by Belle\cite{chi} with a mass of $3872.0\pm0.6\pm0.5$ MeV.
One of its most interesting features is that its energy is within the
error bars of the $D^0D^{0*}$ threshold, $3871.5\pm0.5$ MeV. Considered as a
$c\bar c$ state the most probable assignment would be a $D-$wave,
however most of the quark models predict a somewhat lower mass\cite{chi3}.
Our model does not predict any $q\bar q$ state compatible with this energy.
Due to our tetraquark formalism we can only describe positive parity states\cite{epja}, so
we have studied the $(qc)(\bar q\bar c)$ with $J^P$=$1^+$. We have obtained
3455 MeV for the $I=1$ and 3786 MeV for the $I=0$, both too light to be
identified with the $X(3872)$. We have tried the same approaches
than in the previous case, but all of them would predict an state with an energy
below 3.4 GeV which has not been observed. This seems to
indicate that the $X(3872)$ cannot be described as a tetraquark with $J^P=1^+$. Let us note
that negative parity tetraquarks are always heavier than those with
positive parity, so they seem to be more suitable candidates to describe this
state.

\section{Light scalar sector}
The light scalar sector cannot be described assuming only a $q\bar q$
structure. Our model predicts a pure light content for the $a_0(980)$, what contradicts some of the observed decays,
and an $f_{0}(600)$ too light. Furthermore the $f_{0}(980)$ and the $\kappa (800)$ cannot be found for any
combination of the parameters of the model. We have focussed our
study in the $a_0(980)$ and the $f_0(980)$ as a tetraquark with a structure
$(qs)(\bar q\bar s)$. We obtain 1167 MeV for the isovector case and 1169 MeV
for the isoscalar state with a quark content consistent with their experimental decays\cite{acha}.
This implies that these states are automatically degenerated if we
consider a tetraquark structure. The coupling with $q \bar q$ states or the inclusion of 
three-body color forces should help to improve these results.
%If a three-body interaction with a strength almost equal to the one considered for the
%$D_{sJ}^*(2317)$ is included we obtain 980 MeV and 983 MeV, in good agreement with the
%experimental data and with a quark content consistent with their experimental decays
%\cite{acha}.

\section*{Acknowledgments}
This work has been partially funded by Ministerio 
de Ciencia y Tecnolog{\'{\i}}a under Contract No. BFM2001-3563, 
by Junta de Castilla y Le\'{o}n under Contract No. SA-104/04.

\end{document}